\documentclass{emulateapj}

\begin{document}

\title{A $\sim$5~M$_\earth$ Super-Earth Orbiting GJ~436?: The Power of
Near-Grazing Transits}

\author{Ignasi Ribas\altaffilmark{1,3}, Andreu Font-Ribera\altaffilmark{1,3},
Jean-Philippe Beaulieu\altaffilmark{2,3}}


\altaffiltext{1}{Institut de Ci\`encies de l'Espai (CSIC-IEEC), Campus UAB, 08193 Bellaterra, Spain}
\altaffiltext{2}{Institut d'Astrophysique de Paris, CNRS (UMR 7095), Universit\'e Pierre \& Marie Curie, Paris, France}
\altaffiltext{3}{The HOLMES collaboration}

\begin{abstract}
Most of the presently identified exoplanets have masses similar to that of 
Jupiter and therefore are assumed to be gaseous objects.  With the 
ever-increasing interest in discovering lower-mass planets, several of the 
so-called super-Earths (1 M$_\earth$$<M<$10~M$_\earth$), which are 
predicted to be rocky, have already been found. Here we report the 
possible discovery of a planet around the M-type star GJ~436 with a 
minimum mass of 4.7$\pm$0.6~M$_\earth$ and a true mass of 
$\sim$5~M$_\earth$, which would make it the least massive planet around a 
main-sequence star found to date. The planet is identified from its 
perturbations on an inner Neptune-mass transiting planet (GJ~436b), by 
pumping eccentricity and producing variations in the orbital inclination. 
Analysis of published radial velocity measurements indeed reveals a 
significant signal corresponding to an orbital period that is very close 
to the 2:1 mean motion resonance with the inner planet. The near-grazing 
nature of the transit makes it extremely sensitive to small changes in the 
inclination. 
\end{abstract}


\section{Introduction}

Hundreds of exoplanets have been discovered over the past decade using a 
variety of techniques, notably precision radial velocities, most of them 
being Jupiter-like. While such planets are extremely useful to understand 
the morphology and evolution of planetary systems and star-planet 
connections, there is an obvious interest in eventually identifying an 
Earth-like object. Model calculations indicate that planets with masses in 
the interval 1~M$_{\earth}$$<M<$10~M$_{\earth}$ are of terrestrial type 
\citep[e.g.,][]{IL04,Vea07}.  Such planets have often been dubbed 
``super-Earths''. In this quest for smaller planets there have already 
been some successful detections of planets that fall in the super-Earth 
domain \citep{Rea05,Bea06,Uea07}, and even their habitability may be 
evaluated \citep{Sea07}.

Besides the direct detection from radial velocities or transits, methods 
for the discovery of low-mass planets as perturbers to other planets have 
been suggested (although without any successful detections yet). These are 
mostly based on monitoring variations in the timing of the transit over a 
relatively long time scale so that perturbations by planets as small as a 
few Earth masses could be detected \citep{M02,S04,HM05,Aea05}.

In this {\em Letter} we make use of yet another method that has allowed 
the possible detection of a super-Earth perturbing the transiting planet 
GJ~436b. In this case, it is not perturbations to the transit mid-time but 
to the overall orbital elements of the inner transiting object that cause 
variations on the transit shape and depth. We show that, for near-grazing 
transits, their duration is strongly sensitive to small changes in the 
orbital inclination.

\section{An Unusual Hot-Neptune Around GJ~436}

The M2.5-dwarf GJ~436 was discovered to host a Neptune-mass planet in a 
2.6-d orbit by \citet{Bea04}. Two properties made this object especially 
interesting, namely its relatively small mass and a surprising non-zero 
eccentricity of about 0.15. Such value of the eccentricity was recently 
confirmed by the analysis of \citet{Mea07}. \citet{Bea04} also obtained 
high-precision photometry to investigate the presence of transits but 
ruled out the possibility of a transit with a depth greater than 0.4\%. 
However, a surprise came with the actual detection of transits from 
GJ~436b with a depth of 0.7\% by \citet{Gea07b}, thus becoming, by far, 
the smallest transiting planet yet detected. A series of studies, mostly 
using Spitzer Space Telescope (SST), have greatly contributed to 
establishing the properties of the planet \citep{Dea07,Gea07a,Dea072,T07} 
and also to strengthen the case for an eccentric orbit by observing the 
occultation event at orbital phase 0.59.

The origin of the high eccentricity of GJ~436b was investigated in detail 
by \citet{Mea07} and \citet{Dea07}. Both studies conclude that the 
circularization timescale ($\sim$10$^8$~yr) is significantly smaller than 
the old age of the system ($\gtrsim$6$\cdot$10$^9$~yr) when assuming 
reasonable values for the planet's tidal dissipation parameter. 
\citet{Mea07} also pointed out the presence of a long-term trend with a 
value of 1.3~m~s$^{-1}$ per year on the systemic radial velocity of 
GJ~436. Thus, the authors investigated the possibility that the 
eccentricity and the long-term velocity trend could be explained from the 
perturbation exerted by an object in a wider orbit without reaching 
conclusive results.

We propose an alternative possibility to explain the eccentricity of 
GJ~436b, namely the perturbation from a relatively small planet in a close 
orbit. We show below that the effects of the perturber on the inner planet 
can excite the eccentricity up to the observed value. But GJ~436b has yet 
another characteristic that makes it different to other transiting planets 
and this is the near-grazing nature of its transit. The impact parameter 
of the transit was found to be about 0.85, which implies an orbital 
inclination of 86$\fdg$3. If the inclination happened to be just 85$\fdg$3 
the planet would not cross the disk of the star. Studies mostly focused on 
triple star systems have pointed out that perturbations may not only 
change the eccentricity but also other orbital properties of the inner 
orbit. When the perturber resides in a non-coplanar orbit this gives rise 
to a modulation in the inclination of the inner object 
\citep[e.g.,][]{S75}. As an example, the triple system scenario has been 
advocated to explain the cessation of eclipses in the binary SS Lac 
\citep{TS00,T01}. In the context of exoplanets, \citet{S94}, \citet{M02}, 
\citet{Lea05} considered the possibility that precession induced by a 
perturbing planet could lead to observable changes in the duration (and 
existence) of the transit of an inner giant planet.

GJ~436b makes an ideal system to find evidence for a perturbing small 
planet, because of the telltale non-zero eccentricity, but also to put 
severe constraints on the properties of the perturber owing to the extreme 
sensitivity of the current configuration to small changes in the orbital 
inclination angle.

\section{A second planet around GJ~436?} \label{seccons}

\subsection{Dynamical study}

A possible explanation to the apparently contradicting results concerning 
the detection of transits is that the orbital inclination has indeed 
changed during the 3.3-year interval between the different photometric 
observations. Calculations show that an orbital inclination 
$\lesssim$86$^{\circ}$ would have made the transit undetectable to 
\citet{Bea04}'s photometric measurements. From these considerations a 
small variation of the inclination angle at a rate of roughly 
$\sim$0$\fdg$1~yr$^{-1}$ could make both the \citet{Bea04} non detection 
and \citet{Gea07b}'s discovery of transits compatible. Note that this is 
only a possible scenario since the photometry of \citeauthor{Bea04} has 
relatively sparse phase coverage. In retrospect, from the currently known 
duration and properties of the transit, 3 measurements from \citet{Bea04} 
should have betrayed the presence of the planet, although with low 
significance.

Assuming this hypothesis, it is reasonable to explore the possibility of a 
perturber that could be responsible for both the relatively large 
eccentricity and the inclination change, while remaining undetected by the 
radial velocity measurements. From this, one can assume that the 
semi-amplitude of the perturber should be below about 4~m~s$^{-1}$, 
implying that its mass should satisfy the following inequality: $M_{\rm p} 
\lesssim 30 \, a^{1/2}$~M$_{\earth}$, with $a$ being the orbital 
semi-major axis in AU. On the other hand, because of the inclination 
change, the variation of the transit duration with time is of about 
$10^{-5}$. Following \citet{M02} we find that the perturber mass should 
satisfy: $M_{\rm p} \gtrsim 3\cdot10^{4} \, a^3$~M$_{\earth}$. From both 
inequalities, there is an allowed range of perturber masses and semi-major 
axes.

\begin{figure}
\epsscale{1.0}
\plotone{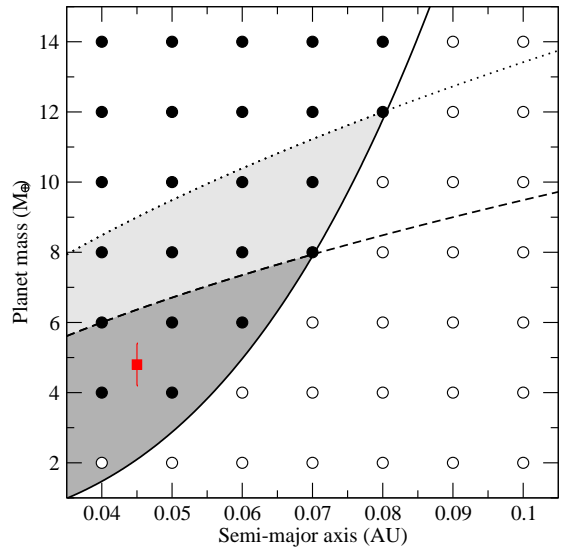}
\caption{Allowed region for a perturbing planet to GJ~436b given the 
observational constraints. The dashed and dotted lines correspond to the 
radial velocity detection limits for inclinations of 90$^\circ$ and 
45$^\circ$, respectively, and the solid line represents the limit from an 
approximate calculation for a perturber (see text). The shaded regions 
illustrate the allowed range of masses and semi-major axes for the 
perturbing object. The circles correspond to a grid search to identify 
cases in which a planetary companion to GJ~436b could be able to induce 
the observed eccentricity and rate of orbital inclination change. Positive 
cases are marked with filled circles while empty circles indicate 
configurations with negative results. The square marks the position 
of the planet detected from the radial velocity analysis.} 
\label{figMerc} 
\end{figure}

For more accurate estimates we carried out direct integrations of the 
equations of motion using the Mercury package \citep{C99}.  We started 
with an inner planet in a circular orbit and with the currently observed 
semi-major axis. Then, we considered different combinations of mass (from 
1 to 14~M$_{\earth}$), semi-major axis (from 0.04 to 0.1 AU), eccentricity 
(from 0.05 to 0.3) and inclination (from 85$^{\circ}$ to 45$^{\circ}$) for 
the perturber. The integrations were performed for a time interval of 
10$^5$ yr to guarantee the stability of the planetary systems.  In Fig. 
\ref{figMerc} we provide an illustration of the region in the mass vs. 
semi-major axis plane where perturbers can meet all constraints, i.e., 
sufficient eccentricity of the inner planet and minimum inclination 
variation rate.

We further explored semi-major axis values at mean-motion resonances 
(MMRs). Location in a MMR can be a stabilizing factor and also 
perturbations can reach their maximum efficiency \citep[e.g.,][]{Aea05}. 
Integrations for semi-major axes corresponding to the following MMRs were 
carried out: 3:2, 5:3, 2:1, 3:1, and 4:1. In all cases, the presence of 
the planet in a MMR increased the stability and, further, perturbing 
planets with smaller masses were able to induce the observed eccentricity 
and orbital inclination change to the inner planet. For the strongest 2:1 
resonance we found a lower limit to the perturbing planet mass of only 
1~M$_{\earth}$ at an extreme eccentricity and relative inclination. For 
the case of a perturbing planet with 3--7~M$_{\earth}$, eccentricity 
values of 0.15--0.20 and initial inclination differences of only 
5--15$^\circ$ were sufficient to explain the observed eccentricity and 
rate of inclination change of the inner planet.

In the analysis we neglected tidal dissipation since we focus on the 
current snapshot of the orbital configuration of the system. The planets 
must be undergoing significant tidal dissipation because of the non-zero 
eccentricity. Our calculation of the dynamical evolution of the system is 
valid to first order because the tidal evolution timescale 
($\sim$10$^{8}$~yr) is long compared with the timescale of the orbital 
perturbations ($\sim$10$^{2-3}$~yr). Other effects have been neglected at 
this stage, which include precession caused by the quadrupole moment of 
the star and by General Relativity. A more detailed dynamical study of the 
system is left for a future paper.

\subsection{Re-analysis of the radial velocity data}

To place more stringent constraints on the perturbing planet we studied 
the available radial velocity observations, covering a time lapse of 7 
years. The orbital solutions of \citet{Bea04} and \citet{Mea07} yield 
radial velocity semi-amplitudes of about 18~m~s$^{-1}$ for GJ~436b, in a 
solution with a rms residual of 4.8~m~s$^{-1}$. We further analyzed the 
data by pre-whitening of the frequencies from the reported Neptune-mass 
planet. The periodogram revealed a relatively strong peak corresponding to 
a period of about 5.2~d. To evaluate the significance of the detection, we 
calculated the false-alarm probability following the Monte Carlo procedure 
in \citet{Bea04}.  From 1000 realizations we conclude that the false-alarm 
probablility of the observed peak is $\sim$20\%, considering that only 
objects with periods between $\sim$4 and $\sim$8 days could be responsible 
for the observed perturbations in the orbital eccentricity and inclination 
of the inner planet. In a recent paper, \citet{Dea072} report 23 
spectroscopic measurements using the HARPS spectrograph of GJ~436. 
Combining these with the data from \citet{Mea07} could provide the needed 
stronger proof of the signal studied here.

\begin{figure}
\epsscale{1.0}
\plotone{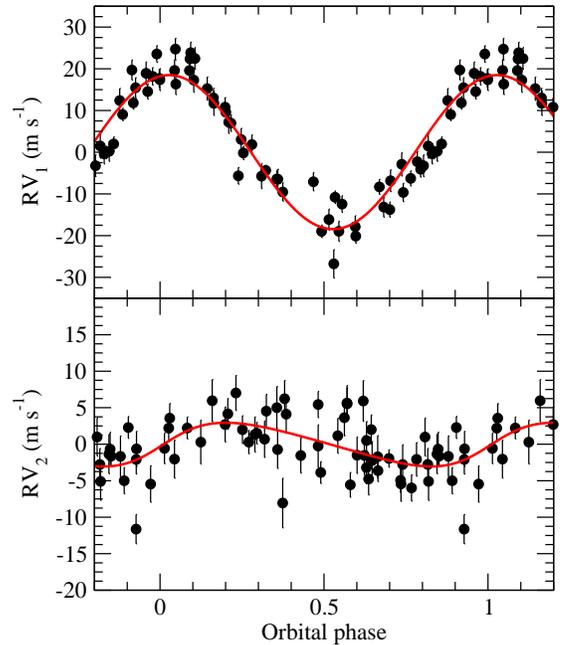}
\caption{Two-planet radial velocity fit to GJ~436. Radial velocity 
observations of GJ~436 \citep{Mea07} were fitted with a model considering 
the orbital motions of two planets combined plus a long term radial 
velocity drift. The panels show the radial velocities (with 1$\sigma$ 
error bars) associated to each respective planet where the contribution 
from the other planet has been removed, together with the best orbital 
fit.} 
\label{figRV} 
\end{figure}

Considering the evidence from the dynamic integrations, the peak in the 
periodogram is sufficiently significant to merit analysis. Starting from 
the 5.2-d period, a simultaneous fit to the orbits of two planets yielded 
the results in Fig. \ref{figRV} and Table \ref{tabRVfit}. In the fit we 
fixed the period of the transiting planet to 2.643913~d, which provides a 
good match to the recent ground-based transit timings, and the 
eccentricity and the argument of periastron to 0.15 and 343$^{\circ}$, 
respectively, for consistency with the observation of the occultation 
event \citep{Dea07,Dea072}. Certainly, the fit to the radial velocities 
should be consistent with our dynamical analysis. This is somewhat 
hampered by two unconstrained parameters, namely the relative inclination 
of the planets and also the true orbital orientation (radial velocities 
are only sensitive to the argument of the periastron). We evaluated a 
range of possible rates of variation in the eccentricities, arguments of 
periastron and orbital periods of the two planets and found them to have 
negligible effects on the fits given the uncertainty of the radial 
velocity measurements and the time baseline. However, we considered a fit 
allowing for a linear change in these three elements for the transiting 
planet and yielded $\dot{e} = 0.03\pm0.02$~yr$^{-1}$, $\dot{w} = 
-1.8\pm1.5^{\circ}$~yr$^{-1}$ and $\dot{P}=-5.7\pm2.0$~s~yr$^{-1}$, all 
with low significance. The orbital elements in Table \ref{tabRVfit} should 
be regarded as osculating elements at the mean epoch of the radial 
velocities. Note that the period given is not sidereal but anomalistic 
(i.e., apparent). From the latter ($P$), the sidereal period ($P_{\rm s}$) 
can be computed as $P_{\rm s} = P (1-\dot{w} P/2\pi)$.

\begin{deluxetable}{lcc}
\tabletypesize{\footnotesize}
\tablecaption{Two-planet fit to the radial velocities. \label{tabRVfit}}
\tablewidth{0pt}
\tablehead{\colhead{Parameter} & \colhead{GJ~436b} & \colhead{GJ~436c}}
\startdata
$P$\tablenotemark{1} (days) & 2.643913 (fixed)     & 5.1859$\pm$0.0013 \\
$T_{\rm peri}$ (HJD)        & 2451551.65$\pm$0.01  & 2451553.2$\pm$0.7\\
$e$                         & 0.15 (fixed)         & 0.2 (fixed)\\
$\omega$ ($^\circ$)         & 343 (fixed)          & 265$\pm$43\\
$K$ (m s$^{-1}$)            & 18.4$\pm$0.4         & 3.0$\pm$0.4 \\
$a$ (AU)                    & 0.0287$\pm$0.0003    & 0.0450$\pm$0.0004 \\
$M \sin i$ (M$_\earth$)     & 23.2$\pm$0.5         & 4.7$\pm$0.6 \\
\hline                      
Radial velocity drift       & \multicolumn{2}{c}{1.1$\pm$0.2} \\
rms (m s$^{-1}$)            & \multicolumn{2}{c}{3.36} \\
${\chi}^2_{\rm red}$        & \multicolumn{2}{c}{3.1} \\
\enddata

\tablenotetext{1}{Anomalistic period.}
\end{deluxetable}

The two-planet fit reduces significantly the rms residuals of the radial 
velocities to 3.4~m~s$^{-1}$. We fixed the orbital eccentricity of the 
second planet because of instabilities in the solution. The adopted 
eccentricity of 0.2 is compatible with our perturbation analysis and 
provides a good fit to the data. The planet's minimum mass is 
4.7$\pm$0.6~M$_\earth$. From the perturbation analysis and planetary 
system formation arguments, the relative inclination of the two planets is 
likely to be below 15$^{\circ}$, and therefore the real mass of the planet 
should be about 5~M$_\earth$ with 12\% uncertainty. While this could not 
be considered an extremely solid detection, its is significant and the 
planet has the correct properties to explain the orbital effects observed 
in GJ~436b.

From the final system configuration, our dynamical calculations predict 
librations of the orbital elements of both planets over different 
timescales and with different amplitudes. Focusing on the inner planet, 
the inclination has a libration amplitude of $\sim$5$^{\circ}$ with a 
period of $\sim$110~yr. Its argument of periastron precesses with a period 
of $\sim$500~yr but also has a libration amplitude of $\sim$20$^{\circ}$ 
over a period of $\sim$70~yr. For comparison, the General Relativistic 
precession period is $\sim$15,000 yr. Finally, the eccentricity has a 
libration amplitude of $\sim$0.1 with a period of $\sim$70~yr.

GJ~436c is the least massive planet known to orbit a main-sequence star 
and only the second bona-fide warm super-Earth together with the 
7.5$\pm$0.7~M$_\earth$ planet around GJ~876 \citep{Rea05}, since only 
minimum masses are available for the planets around Gl~581 \citep{Uea07}. 
Interestingly, GJ~436c is found at an orbital period that is very close to 
the 2:1 MMR with GJ~436b, but not exactly so: $P_c/P_b = 1.9614\pm0.0005$.  
This small but significant departure may be a consequence of the tidal 
evolution of the system.

\section{Prospects for Confirmation}

\begin{figure}
\epsscale{1.0}
\plotone{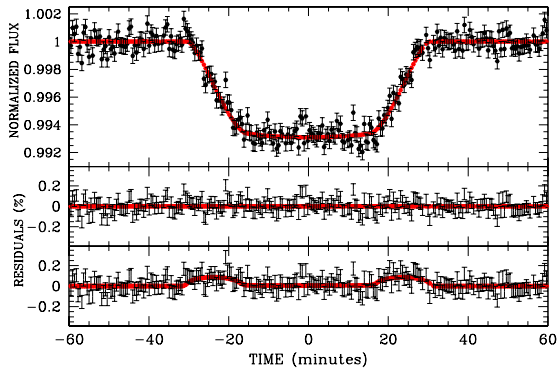}
\caption{Effect of a +0$\fdg$1 change in the orbital inclination of 
GJ~436b. The upper panel shows the best fit to the 8~$\mu$m transit data 
from SST \citep{Gea07a}. The middle panel shows the residuals from the 
fit, which result in a $\chi^2$ value of 486 with 355 degrees of freedom. 
The bottom panel illustrates the residuals of the fit when assuming an 
orbital inclination higher by +0$\fdg$1, with a $\chi^2$ increasing to 
578.}
\label{figtr}
\end{figure}

Because of the inclination change, the effects of the perturbing planet 
will become evident in high-precision transit photometry collected during 
2008. We carried out tests to assess the capability to detect small 
changes in the inclination using real SST IRAC primary transit 
observations. We adopted a non-linear limb darkening law model with 4 
coefficients and followed the procedure described in \citet{MA02} to fit 
the transit light curve. Then, we fitted the same dataset with models for 
an inclination larger by 0$\fdg$1. Such expected inclination change will 
increase the transit duration in $\sim$2 minutes making it detectable from 
SST at reasonably high confidence level. Fig. \ref{figtr} shows the best 
fit to the 8~$\mu$m SST data and the residuals.

\section{Conclusions}

The proposed existence of GJ~436c is based on three independent pieces of 
evidence: {\em a)} An orbital inclination change supported by the lack of 
transit detection in 2004 and the grazing transit observed in 2007; {\em 
b)} an inner planet with significant orbital eccentricity and a tidal 
dissipation timescale much smaller than the age of the system; and {\em 
c)} a low-amplitude radial velocity signal that is fully consistent with a 
planet in the 2:1 mean motion resonance with the inner planet. A key 
element of our study is a dynamical investigation that explains the 
observed effects {\em a)} and {\em b)}, and predicts the existence of a 
perturbing planet with constraints on its mass and semi-major axis. 
Reanalysis of the radial velocity data indeed reveals such planet closely 
matching the predicted properties. In other words, the eccentricity of the 
orbit clearly reveals the existence of a perturbing planet, and a grazing 
transit will be most sensitive to even mild non-coplanarity between the 
two objects. We find such possible inclination change and we identify a 
radial velocity signal matching the predictions of our dynamical 
integrations.

The resulting planet system around GJ 436 strengthens the trend of a 
relatively large number of hot-Neptunes and super-Earths around stars of 
low mass \citep{Bea07}. The presence of a long-term radial velocity trend 
of $\sim$1~m~s$^{-1}$ per year could still be indicative of further 
planets in the system with wider orbits. Indeed, the system around GJ~436 
shows striking resemblances to that around the M-type star Gl~581 
\citep{Uea07}, and thus its planets may experience changes in the orbital 
elements, perhaps eventually undergoing transits in spite of a previously 
null result \citep{LMea06}. Our study provides yet another illustration of 
the variety of exoplanet systems and highlights the potential for complex 
dynamical histories that imply sizeable variations of the planets' orbital 
elements over timescales of decades.

The method of using near-grazing transits should be of special interest to 
discover small planets. Even for mild non-coplanarity, objects as small as 
a few M$_{\earth}$ can lead to moderately high values of the induced 
eccentricity and to orbital inclination changes in the order of tenths of 
degrees per year. This will be especially effective for transit search 
missions from space, which could overcome the lower detection probability 
of near-grazing transits, and push the detection limits to even lower-mass 
objects than those responsible for the transit events.

\acknowledgements

We are indebted to G. Tinetti for her continuous help and support in the 
completion of this work. We are grateful to J. Miralda-Escud\'e for 
fruitful discussions. We thank the referee, G. Laughlin, for constructive 
criticism. I.R. and A.F.-R. acknowledge financial support from the Spanish 
MEC through grant AyA2006-15623-C02-02. A.F.-R. thanks the Spanish CSIC 
for support via a research fellowship. We acknowledge financial support 
from the ANR HOLMES.

\end{document}